\documentclass{appolb}
\usepackage{epsfig}
\usepackage{amssymb}
\begin{document}
\title{Inclusive and diffractive deep inelastic scattering in high-energy QCD
\thanks{Presented at the XLVIth Cracow School of Theoretical Physics, Zakopane, 
Poland, May 27-June 5 2006}}
\author{Cyrille Marquet
\address{Service de Physique Th\'eorique, CEA/Saclay\\
91191 Gif-sur-Yvette cedex, France\\email: marquet@spht.saclay.cea.fr}}

\maketitle
\begin{abstract}

In the context of both inclusive and diffractive deep inelastic scattering,
we derive the first phenomenological consequences of the inclusion of Pomeron 
loops in the QCD evolution equations towards high energy. We discuss the 
transition between the well-known geometric scaling regime and the new diffusive 
scaling, that emerges for sufficiently high energies and up to very large values 
of $Q^2,$ well above the proton saturation momentum.
\end{abstract}


\section{The Good-and-Walker picture in high-energy QCD}

The Good-and-Walker picture~\cite{goodwalker} of diffraction was originally 
meant to describe soft diffraction. They express an hadronic projectile 
$|P\rangle\!=\!\sum_n c_n|e_n\rangle$ in terms of hypothetic eigenstates of the 
interaction with the target $|e_n\rangle,$ that can only scatter elastically:
$\hat{S}|e_n\rangle\!=\!(1\!-\!T_n)|e_n\rangle.$ The total, elastic and 
diffractive cross-sections are then easily obtained:
\begin{equation}
\sigma_{tot}=2\sum_n c_n^2 T_n\hspace{0.5cm}
\sigma_{el}=\Big[\sum_n c_n^2 T_n\Big]^2\hspace{0.5cm}
\sigma_{diff}=\sum_n c_n^2 T_n^2\ .\label{gaw}
\end{equation}
 
It turns out that in the high energy limit, there exists a basis of eigenstates 
of the large$-N_c$ QCD $S-$matrix: sets of quark-antiquark color dipoles 
$|e_n\rangle\!=\!|d(r_1),\dots,d(r_n)\rangle$ caracterized by their transverse 
sizes $r_i.$ In the context of deep inelastic scattering (DIS), we also know the 
coefficients $c_n$ to express the virtual photon in the dipole basis. For 
instance, the equivalent of $c_1^2$ for the one-dipole state is the well-known 
photon wavefunction $\phi(r,Q^2)$ where $Q^2$ is the virtuality of the photon. 

This realization of the Good-and-Walker picture allows to write down exact 
(within the high-energy and large$-N_c$ limits) factorization 
formulae~\cite{difscal} for the total, elastic and diffractive cross-sections in 
DIS. They are expressed in terms of elastic scattering amplitudes of dipoles off 
the target proton $\left\langle T_n(\{r_i\})\right\rangle_Y,$ where the average 
$\langle\ .\ \rangle_Y$ is an average over the proton wavefunction that gives 
the energy dependence to the cross-sections ($Y\!\sim\!\log(s)$ is the 
rapidity). 

Formulae are similar to (\ref{gaw}) with extra integrations over the dipoles 
transverse coordinates. For instance, denoting the total rapidity $Y$ and the 
minimal rapidity gap $Y_g,$ the diffractive cross-section reads~\cite{difscal}
\begin{equation}
\sigma_{diff}(Y,Y_g,Q^2)=\sum_n\int dr_1\cdots dr_n\ 
c_n^2(\{r_i\},Q^2,Y\!-\!Y_g)\  \left\langle T_n(\{r_i\})\right\rangle_{Y_g}^2\ 
.\label{diff}
\end{equation}
This factorization is represented in Fig.1. Besides the $Q^2$ dependence, the 
probabilities to express the virtual photon in the dipole basis $c^2_n$ also 
depend on $Y\!-\!Y_g.$ Starting with the initial condition 
$c_n^2(\{r_i\},Q^2,0)\!=\!\delta_{1n}\phi(r,Q^2),$ the probabilities can be 
obtained from the high-energy QCD rapidity evolution. Finally, the scattering 
amplitude of the n-dipole state $T_n(\{r_i\})$ is given by
\begin{equation}
T_n(\{r_i\})=1-\prod_{i=1}^n(1-T(r_i))
\end{equation}
where $T(r)\equiv T_1(r)$ is the scattering amplitude of the one-dipole state. 
We are therefore led to study the rapidity evolution of objects such as 
$\left\langle T(r_1)\dots T(r_n)\right\rangle_Y.$

\begin{figure}[t]
\centerline{\epsfxsize=10cm\epsfbox{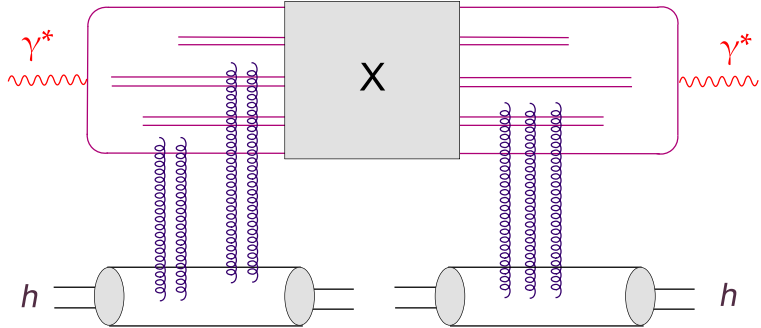}}
\caption{Representation of the factorization formula (\ref{diff}) for the 
diffractive cross-section in DIS. The virtual photon is decomposed into dipoles
which interact elastically with the target hadron. The rapidity gap is $Y_g$ and 
the final state $X$ is made of particles produced over a rapidity interval 
$Y-Y_g.$}
\end{figure}

\section{The geometric and diffusive scaling regimes}

Within the high-energy and large$-N_c$ limits, the dipole amplitudes are 
obtained from the Pomeron-loop equation~\cite{ploop} (see Fig.2) derived in the 
leading logarithmic approximation in QCD (see also~\cite{greg}). This is a 
Langevin equation which exhibits the stochastic nature~\cite{fluc} of 
high-energy scattering processes in QCD (see also~\cite{steph}). Its solution is 
an event-by-event dipole scattering amplitude function of 
$\rho\!=\!-\log(r^2Q_0^2)$ and $Y$ ($Q_0$ is a scale provided by the initial 
condition). It is characterized by a saturation scale $Q_s$ which is a random 
variable whose logarithm is distributed according to a Gaussian probability 
law~\cite{prob}. The average value is $\log(\bar{Q}_s^2/Q_0^2)\!=\!\lambda Y$ 
and the variance is $\sigma^2\!=\!DY$ (see Fig.3, left plot). The dispersion 
coefficient 
$D$ allows to distinguish between two energy regimes: the geometric scaling 
regime ($DY\!\ll\!1$) and diffusive scaling regime ($DY\!\gg\!1$).

\begin{figure}[t]
\centerline{\epsfxsize=6cm\epsfbox{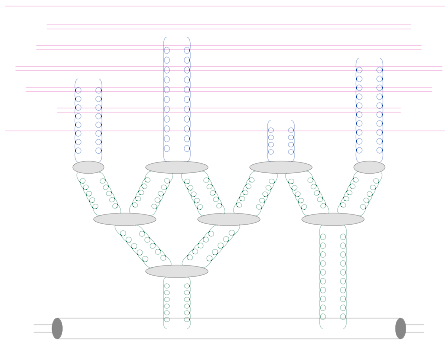}}
\caption{Elastic scattering of a multiple-dipole state off the target hadron.
Shown is a typical contribution included in the Pomeron-loop equation for the 
amplitudes $\left\langle T(r_1)\dots T(r_n)\right\rangle_Y.$ In Fig.1, the 
vertical gluon lines representing the interaction with the hadron actually stand 
for this type of diagram.}
\end{figure}

The following results for the averaged amplitude will be needed to derive the 
implications for inclusive and diffractive DIS:
\begin{eqnarray}
\left\langle T(r_1)\dots T(r_n)\right\rangle_Y&\stackrel{Y\ll1/D}{=}&
\left\langle T(r_1)\right\rangle_Y\dots\left\langle T(r_n)\right\rangle_Y\ ,\\
\left\langle T(r_1)\dots T(r_n)\right\rangle_Y&\stackrel{Y\gg1/D}{=}&
\left\langle T(r_<)\right\rangle_Y,\hspace{0.3cm}r_<=\min(r_1,\dots,r_n)\ .
\end{eqnarray}
All the scattering amplitudes are expressed in terms of $\langle T(r)\rangle_Y,$ 
the amplitude for a single dipole which features the following scaling 
behaviors:
\begin{eqnarray}
\left\langle 
T(r)\right\rangle_Y\stackrel{Y\ll1/D}{\equiv}T_{gs}(r,Y)&=&\displaystyle
T\left(r^2\bar{Q}_s^2(Y)\right)\ ,\label{gs}\\
\left\langle 
T(r)\right\rangle_Y\stackrel{Y\gg1/D}{\equiv}T_{ds}(r,Y)&=&\displaystyle
T\left(\frac{\log(r^2\bar{Q}_s^2(Y))}{\sqrt{DY}}\right)\ .\label{ds}
\end{eqnarray}
In the saturation region $r\bar{Q}_s\!>\!1,$ $\left\langle 
T(r)\right\rangle_Y\!=\!1.$ As the dipole size $r$ decreases, $\left\langle 
T(r)\right\rangle_Y$ decreases towards the weak-scattering regime, following the 
scaling laws (\ref{gs}) or (\ref{ds}), depending on the value of $DY$ (see 
Fig.3, right plot). In the geometric scaling regime ($DY\!\ll\!1$), the 
dispersion of the events is negligible and the averaged amplitude obeys 
(\ref{gs}). In the diffusive scaling regime ($DY\!\gg\!1$), the dispersion of 
the events is important, resulting in the behavior (\ref{ds}).

\begin{figure}[t]
\begin{minipage}[t]{45mm}
\centerline{\epsfxsize=4.5cm\epsfbox{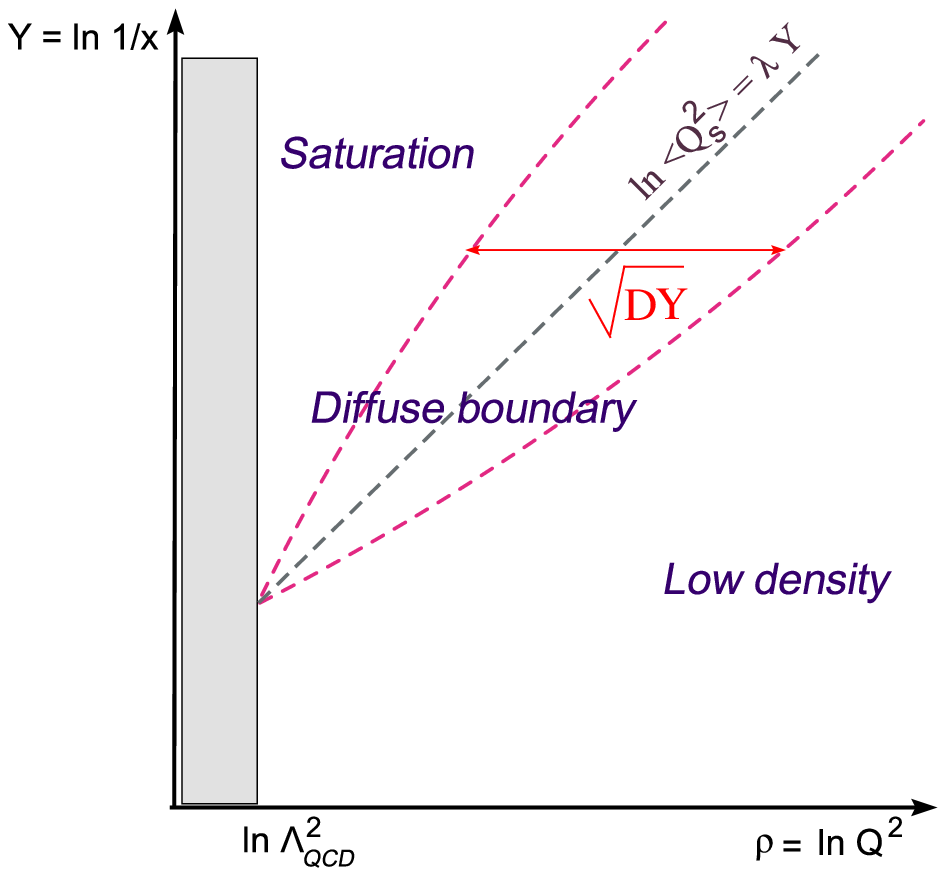}}
\end{minipage}
\hspace{\fill}
\begin{minipage}[t]{75mm}
\vspace{-4.3cm}\centerline{\epsfxsize=4.3cm\rotatebox{-90}{\epsfbox{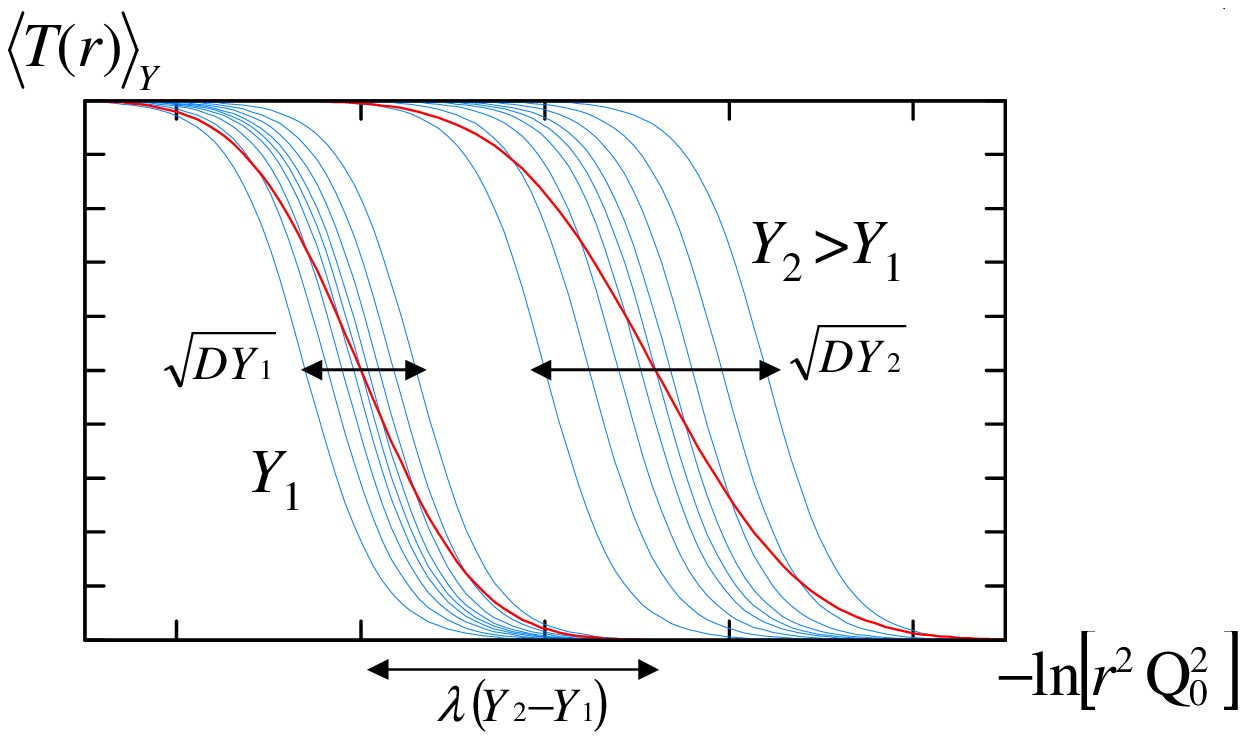}}}
\end{minipage}
\caption{Left plot: a diagram representing the stochastic saturation line in the 
$(\rho,Y)$ plane, the diffusive saturation boundary is generated by the 
evolution. Right plot: different realizations of the event-by-event scattering 
amplitude (gray curves) and the resulting averaged physical amplitude $\langle 
T(r)\rangle$ (black curve) as a function of $\rho$ for two different values of 
$Y$ in the diffusive scaling regime.}
\end{figure}

\section{Implications for inclusive and diffractive DIS}

We shall concentrate on the diffusive scaling regime, in which the dipole 
scattering amplitude can be written as follows~\cite{errorf} for 
$-\log(r^2\bar{Q}_s^2(Y))\!\ll\!DY:$
\begin{equation}
T_{ds}(r,Y)=\frac12 
Erfc\left(-\frac{\log(r^2\bar{Q}_s^2(Y))}{\sqrt{DY}}\right)\ .
\label{erfc}\end{equation}
From this, one obtains the following analytic estimates \cite{difscal} for the 
$\gamma^*\!-\!p$ total cross-section in DIS and for the diffractive 
cross-section integrated over the rapidity gap size (at fixed $Y\!=\!\log(1/x)$) 
from $Y_g\!=\!\log(1/\beta_<)$ to $Y:$
\begin{eqnarray}
\frac{d\sigma_{tot}}{d^2b}(x,Q^2)&=&\frac{N_c\alpha_{em}}{12\pi^2}\sum_f e_f^2
\sqrt{\pi D\log(1/x)}\ \frac{e^{-Z^2}}{Z^2}\ ,\\
\frac{d\sigma_{diff}}{d^2b}(x,Q^2,\beta_<)&=&\frac{N_c\alpha_{em}}{48\pi^2}
\sum_f e_f^2 \sqrt{D\log(1/x)}\ \frac{e^{-2Z^2}}{Z^3}\ .
\end{eqnarray}
The variable $Z$ is reminiscent of the scaling variable of the dipole amplitude: 
\begin{equation}
Z=\frac{\log(Q^2/\bar{Q}_s^2(x))}{\sqrt{D\log(1/x)}}\ .
\end{equation}
It shows that in the diffusive scaling regime, both inclusive and diffractive 
scattering are dominated by small dipole sizes $r\!\sim\!1/Q.$ Also the 
cross-sections do not feature any Pomeron-like (power-law type) increase with 
the energy and the diffractive cross-section (which does not depend on 
$\beta_<$) is dominated by the scattering of the quark-antiquark ($q\bar q$) 
component, corresponding to rapidity gaps close to $Y.$ These features a priori 
expected in the saturation regime ($Q^2<\bar{Q}_s^2$) are valid up to values of 
$Q^2$ much bigger than $\bar{Q}_s^2:$ in the whole diffusive scaling regime for 
$\log(Q^2/\bar{Q}_s^2(Y))\!\ll\!DY$ (see Fig.4).

\begin{figure}[t]
\centerline{\epsfxsize=8cm\epsfbox{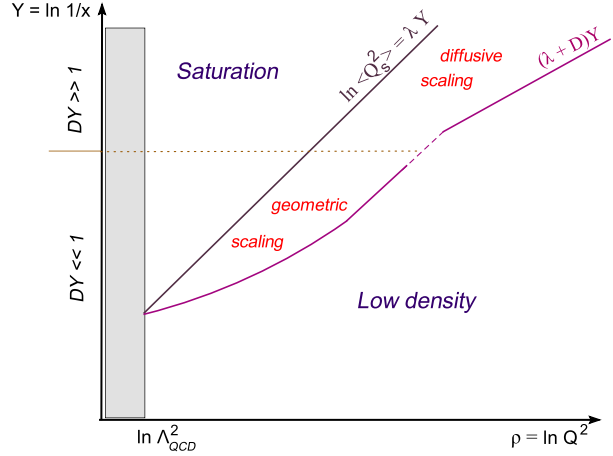}}
\caption{A phase diagram for the high-energy limit of inclusive and diffractive 
DIS in QCD. Shown are the average saturation line and the approximate boundaries 
of the scaling regions at large values of $\rho\!\sim\!\ln Q^2.$ With increasing 
$Y,$ there is a gradual transition from geometric scaling at intermediate 
energies to diffusive scaling at very high energies.}
\end{figure}

The inclusive cross-section and the $q\bar q$ contribution to the diffractive 
cross-section are obtained from the dipole amplitude $\langle T(r)\rangle_Y$ in 
the following way:
\begin{eqnarray}
\frac{d\sigma_{tot}}{d^2b}&=&2\pi\int dr^2 \Phi(r,Q^2)\langle T(r)\rangle_Y\ , 
\label{dis}\\
\frac{d\sigma_{diff}}{d^2b}&=&\pi\int dr^2 \Phi(r,Q^2)\langle T(r)\rangle^2_Y\ .
\label{ddis}
\end{eqnarray}
In order to better exhibit the dominance of small dipole sizes $r\!\sim\!1/Q$, 
we represent in Fig.5 the integrands of (\ref{dis}) and (\ref{ddis})  as a 
function of the dipole size $r.$ Keeping $Q/\bar{Q}_s\!=\!10$ fixed, we use 
(\ref{erfc}) in the diffusive scaling regime and 
$T_{gs}(r,Y)\!=\!1\!-\!e^{-r^2\bar{Q}_s^2(Y)/4}$ in the geometric scaling 
regime. 

The difference between the two regimes is striking. In the geometric scaling 
regime, the total cross-section is dominated by semi-hard sizes 
($1/Q\!<\!r\!<\!1/\bar{Q}_s$) while the diffractive cross-section is dominated 
by inverse dipole sizes of the order of the hardest infrared cutoff in the 
problem: the average saturation scale $\bar{Q}_s.$ In the diffusive scaling 
regime, both inclusive and diffractive scattering are dominated by inverse 
dipole sizes of the order of the hardest infrared cutoff in the problem: the 
hardest fluctuation of the saturation scale, which is as large as $Q.$

In the diffusive scaling regime, up to values of $Q^2$ much bigger than the 
saturation scale $\bar{Q}_s^2$, cross-sections are dominated by rare events in 
which the photon hits a black spot that he sees at saturation at the scale 
$Q^2.$ In average the scattering is weak, but saturation is the only relevant 
physics.

\begin{figure}[t]
\centerline{\epsfxsize=6.5cm\rotatebox{-90}{\epsfbox{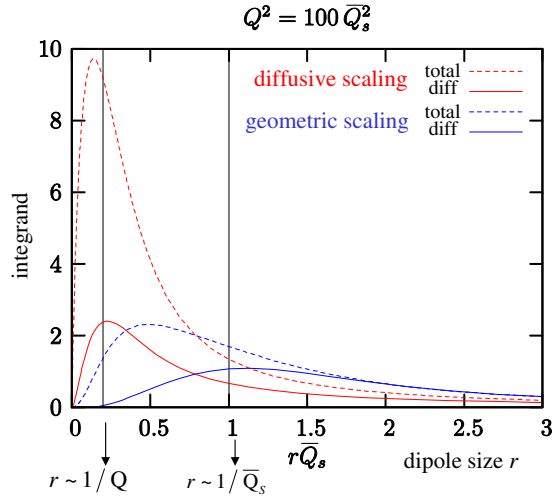}}}
\caption{The integrands of (\ref{dis}) and (\ref{ddis})  plotted as a function 
of $r\bar{Q}_s$ (with $Q/\bar{Q}_s\!=\!10$ fixed) and computed with
two 
expressions for the dipole amplitude: in the geometric and diffusive scaling 
regimes.}
\end{figure}

\section*{Acknowledgments}

The work described here was done in collaboration with Yoshitaka Hatta, Edmond 
Iancu, Gr\'egory Soyez and Dionysis Triantafyllopoulos. I would like to thank 
the organizers and especially Michal Praszalowicz for giving me the opportunity 
to present this work at the school.

\end{document}